\documentclass[12pt]{iopart}
\usepackage{graphicx}% Include figure files

\begin{document}

\title[ ]{Device fabrication and transport measurements of FinFETs built with $^{28}$Si SOI wafers towards donor qubits in silicon}

\author{Cheuk Chi Lo$^{1,3}$, Arun Persaud$^3$, Scott Dhuey$^2$, Deirdre Olynick$^2$, Ferenc Borondics$^2$, Michael C. Martin$^3$, Hans A. Bechtel$^3$, Jeffrey Bokor$^{1,2}$ and Thomas Schenkel$^3$}

\address{$^1$ Department of Electrical Engineering and Computer Sciences, University of California, Berkeley, CA 94720, USA}
\address{$^2$ The Molecular Foundry, E.O. Lawrence Berkeley National Laboratory, Berkeley, CA 94720, USA}
\address{$^3$ Accelerator and Fusion Research Division, E.O. Lawrence Berkeley National Laboratory, Berkeley, CA 94720, USA}
\address{$^4$ Advanced Light Source, E.O. Lawrence Berkeley National Laboratory, Berkeley, CA 94720, USA}

\ead{cclo@eecs.berkeley.edu} 

\begin{abstract}
We report fabrication of transistors in a FinFET geometry using isotopically purified silicon-28 -on-insulator (28-SOI) substrates.  Donor electron spin coherence in natural silicon is limited by spectral diffusion due to the residual $^{29}$Si nuclear spin bath, making isotopically enriched nuclear spin-free $^{28}$Si substrates a promising candidate for forming spin quantum bit devices.   The FinFET architecture is fully compatible with single-ion implant detection for donor-based qubits, and the donor spin-state readout through electrical detection of spin resonance.  We describe device processing steps and discuss results on electrical transport measurements at 0.3 K. 
\end{abstract}


%\submitto{\SST} 

\maketitle

\section{Introduction}
Silicon based quantum computation with donor qubits has attracted much attention since its original proposal by Kane \cite{kane98}. The prevalence and maturity of silicon processing technologies offer a convenient platform for pursuing silicon-based quantum computation schemes. More importantly, it has been long known that donor electron and donor nuclei spins have extraordinary long spin-coherence times at cryogenic temperatures in isotopically purified $^{28}$Si substrates \cite{tyryshkin03,morton08}, with the long spin lifetime being attributed to the weak spin-orbit interaction in the crystal lattice and the nuclear spin-free matrix. However, many challenges have to be met in order to demonstrate donor spin qubits, including the fabrication of a readout device doped with single donors \cite{simmons05,shen02,hopf05,batra07,shinada08} , and the implementation of a suitable single spin-state measurement.

Single-ion implantation is a technique that enables single atom placement for donor qubit device formation.  Early single-ion implants were achieved by sensitive detection of secondary electrons from single-ion implantation events \cite{shinada98}.  More recently, electrical detection of single-ion impacts have been achieved with the use of {\it p-i-n} photodetector-like structures \cite{hopf05,jamieson05} as well as with metal-oxide-semiconductor (MOS) field-effect transistor (FET) architectures \cite{batra07,shinada08}. In our earlier work \cite{batra07}, conventional planer FETs were used with an aperture formed in the middle of the gate, allowing implanted ions to travel through. While the experiment serves as a proof-of-concept single-ion detection scheme using FETs, the planer device structure is not ideal as the gate has to be partially removed in order for the single-ion implantation event to take place. The partial gate removal potentially damages the silicon channel and introduces instability issues into the device. To circumvent this problem, we explore the possibility of utilizing the non-planer device architecture of the FinFET. The FinFET architecture allows both single-ion implantation detection and spin-state readout capabilities via electrically detected magnetic resonance (EDMR) \cite{desousa08,sarovar08}. Recently, we reported studies of ion implant detection and ion impact mapping using FinFETs \cite{weis08}. In this paper, we describe our device design strategies, the device fabrication process and first low temperature transport measurements of FinFETs formed in isotopically purified silicon-28 -on-insulator (28-SOI).

\section{Device design}
\begin{figure}
\includegraphics[width=4in]{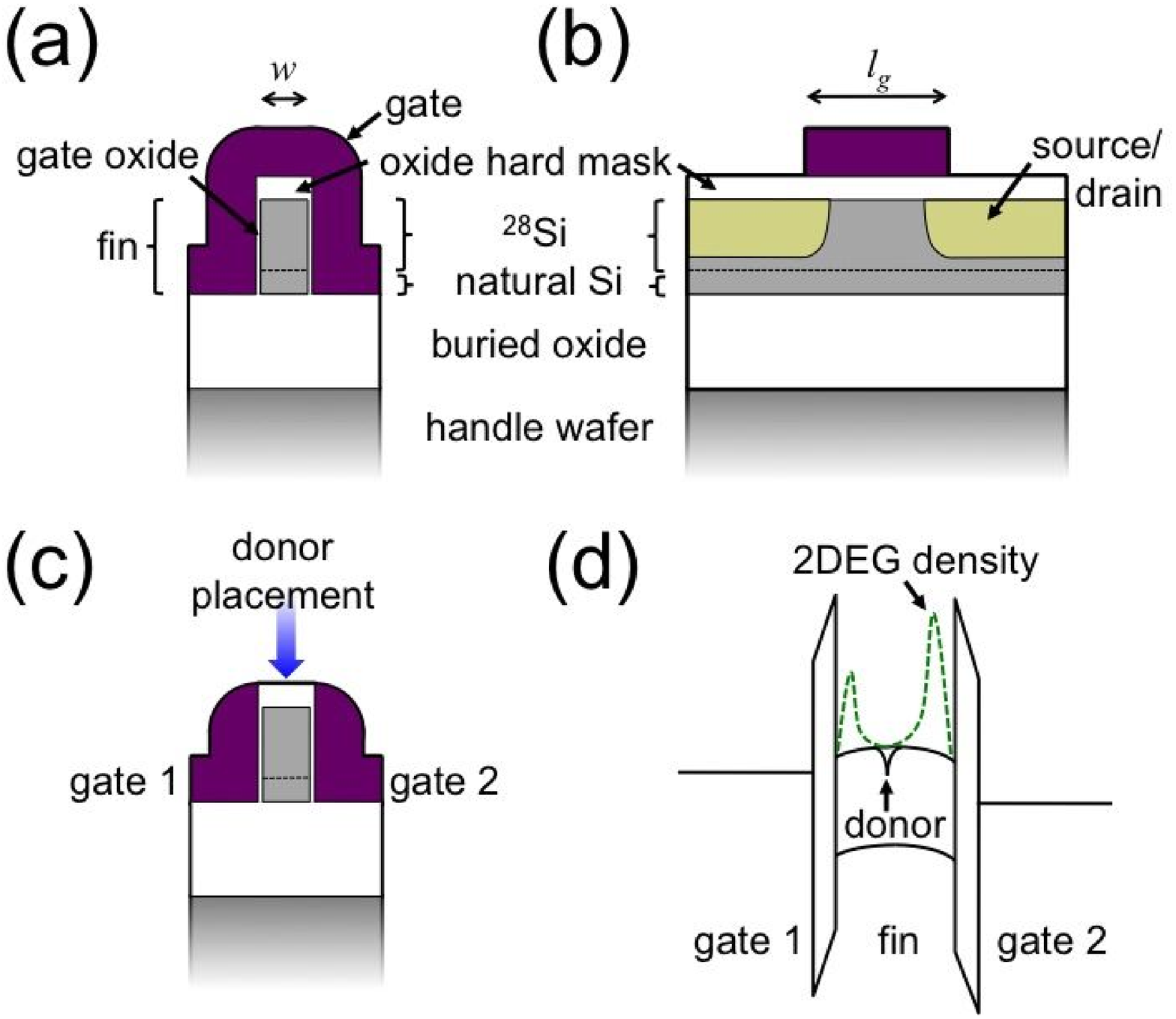}
\caption{Schematics of the FinFETs formed with 28-SOI wafers. (a) The SOI fin consists of 100 nm thick natural silicon and 150 nm of epitaxially grown $^{28}$Si with $>$ 99.9$\%$ purification.  (b) Cross-section of the $^{28}$Si FinFET along the source/drain direction. (c) FinFET with top portion of gate removed, allowing single-ion implantation to take place. (d) Schematics of the energy-band diagram and 2DEG densities of a split-gate FinFET as shown in (c). A split-gate device allows better control of the 2DEG-donor interaction by applying independent biases to the gates. $l_g$ and $w$ indicate the physical gate length and fin width, respectively.
\label{schematic}}
\end{figure}

While FinFETs were invented as an end-of-roadmap CMOS technology with extremely scaled dimensions ($<$ 50 nm gate length) \cite{hisamoto00}, they also provide a convenient architecture to achieve single-ion implantation and spin-state detection with the same device for realizing donor qubits. Several factors with the objective of silicon quantum computation in mind establishes the baseline for the modified device design: Shallow donors in silicon have Bohr radii of approximately 2 nm, hence the donor electron wavefunction can extend to $>$10 nm from the position of the donor nuclei. While donor electrons in bulk silicon have extraordinarily long spin coherence lifetimes, the presence of the oxide-silicon interfaces degrade the spin coherence time \cite{schenkel06}. Hence, donors should be placed at least 15 nm away from the surface so that the donor electron interacts minimally with the interface spin-noise sources \cite{desousa07}, but at the same time kept shallow enough to interact with the gate induced two-dimensional electron gas (2DEG) that is used for spin-state readout detection in an EDMR experiment. In the case of FinFETs, the oxide interface is present on both sides of the donor (Fig.~\ref{schematic}a, b), hence we limit the minimal width of the fins ($ w$) to approximately 40 nm. Since the 2DEG is induced on the side-walls of the fin in a FinFET, the top portion of the gate can be removed for single-ion implantation without affecting the gating ability of the side gates (Fig.~\ref{schematic}c). In additional, if separate contact leads are designed for the FinFet, the two side gates can then be biased independently for optimal 2DEG-donor interaction with the split-gate geometry (Fig.~\ref{schematic}d). 

The donor atom that would serve as the qubit has to be isolated from other impurity atoms. However, in a conventional CMOS process, source and drain regions are degenerately doped and some of the dopants might straggle during ion implantation or diffuse into the channel, as is the case in recent single dopant transport measurements in tri-gate FETs \cite{sellier06}. Hence, the gate lengths ($l_g$) for spin-state readout FinFETs are designed to be $>$ 250 nm as TCAD simultions indicate that the channel region will not be accidentally doped by the source/drain dopants for devices with such long gate lengths, given the processing conditions used. In addition, we use a different donor species in the channel donor implant (antimony) and for the source/drain degenerate implants (arsenic), and dopant species identification can be achieved through spectroscopic transport measurements such as EDMR\cite{ghosh92,lo07}.

Silicon-on-insulator (SOI) wafers are used for FinFET fabrication. However, for spin qubits the presence of $^{29}$Si isotopes in natural silicon reduces the spin coherence lifetime dramatically due to spectral diffusion \cite{tyryshkin03}, and nuclear spin free materials are critical for spin qubit device development \cite{sailer09}.  The lack of commercially available isotopically enriched $^{28}$Si SOI wafers led us to adopt a hybrid approach: $^{28}$Si is epitaxially grown on a thin natural silicon SOI layer.  When donors are implanted into the device, we use sufficiently low implantation energies so that the donor electrons reside only in the isotopically purified $^{28}$Si environment. After single-ion implantation is detected electrically, the device has to be annealed at a high temperature for dopant activation. Hence we use only tungsten for the metal layer of the device. The complete process flow is described in detail in the next section.

\section{Device fabrication}
\begin{figure}
\includegraphics[width=4in]{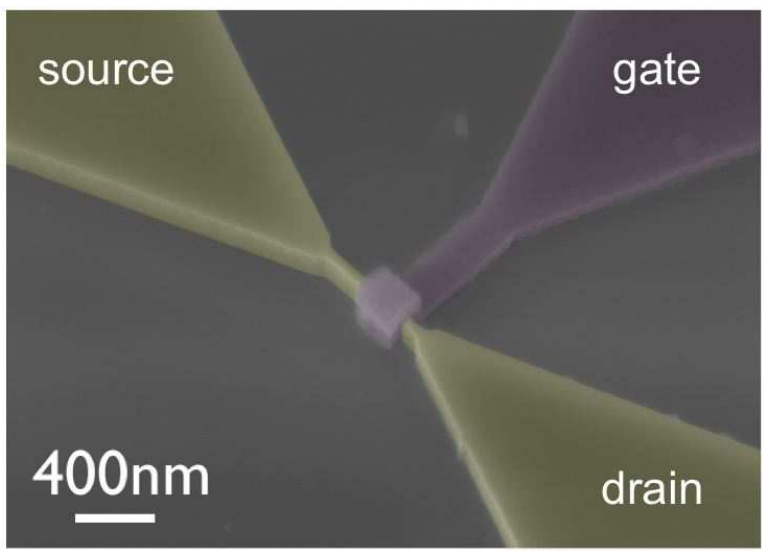}
\caption{False-colour SEM micrograph of a 28-SOI FinFET close to completion. The micrograph was taken with the sample tilted at 30$^o$. The device shown has a gate length of $l_g$ = 280 nm and fin width of $w$ = 80 nm, similar to the one measured (see text).
\label{SEM}}
\end{figure}

\begin{figure}
\includegraphics[width=4in]{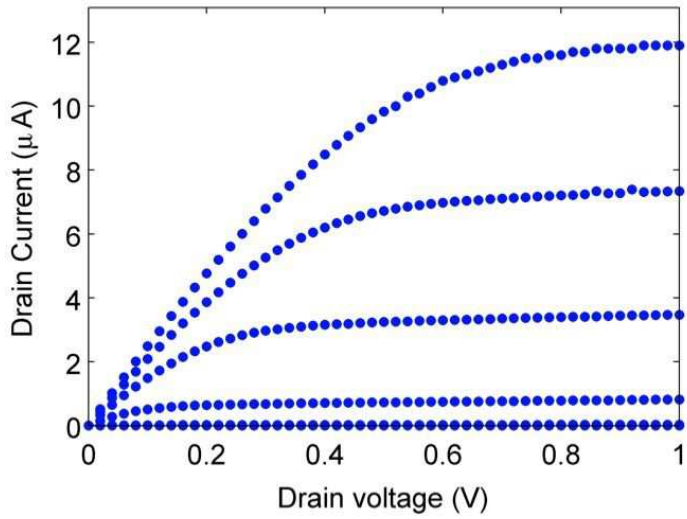}
\caption{Typical room temperature $I_d-V_d$ characteristics of fabricated FinFETs ($l_g$ = 280 nm, $w$ = 80 nm). The gate voltage is stepped from - 0.2 V to + 0.6 V in 0.2 V increments. 
\label{idvd}}
\end{figure}

The starting substrates were silicon-on-insulator (SOI) wafers with natural isotope compositions from Soitech.  The top $<100>$ silicon layer thickness was 100 nm and the box thickness was 200 nm. 150 nm of isotopically enriched $^{28}$Si ($>$99.9\%) was epitaxially grown on the original SOI wafers, bringing the total top layer silicon thickness to 250 nm. A 100 nm thick low-temperature chemical-vapor deposited (CVD) silicon oxide layer was deposited to serve as a hard-mark for the fins. Electron beam lithography was used to pattern the fin patterns along with the source/drain contact pads. The oxide hard-mask and the SOI layer were then etched by reactive-ion etching to define the fins. After the fin etch, a 3 nm thick sacrificial dry oxide was grown and removed in dilute hydrofluoric acid to smoothen the SOI sidewall. Following sidewall smoothing, a 10 nm thick dry oxide was grown, and 140 nm thick {\it in-situ} phosphorus-doped poly-crystalline silicon was deposited as the gate material. A second electron beam lithography step was carried out to define the gate patterns, and the poly-crystalline silicon gate was patterned by another reactive-ion etching step. The wafers then received an arsenic implant with a dose of $2\times10^{15}$/cm$^2$ at 25 keV to create self-aligned degenerately doped source/drain regions. Dopant activation was achieved by rapid thermal annealing. A protective CVD oxide, tungsten metallization, and forming gas anneal completes the basic device fabrication process. The thermal budget of the full device fabrication was designed with consideration of self diffusion of $^{29}$Si from the 100 nm natural silicon layer into the $^{28}$Si epitaxial layer.  Control measurements by electron spin resonance at 9 K showed linewidths of implanted $^{121}$Sb donors in the 28-SOI layers to be 0.2 G \cite{tyryshkin08}, while the linewidth in natural siliocn is about 4 G \cite{tyryshkin03}, confirming the integrity of the isotope enriched $^{28}$Si layer.  From literature values of diffusivities \cite{bracht98}, the expected self-diffusion of $^{29}$Si under the thermal budgets used for device fabrication is only about 4 nm. A SEM micrograph of a fabricated FinFET is shown in Fig.~\ref{SEM}, prior to the final CVD oxide deposition and metallization steps. A typical room temperature $I_d-V_d$ measurement of the devices is shown in Fig.~\ref{idvd}. 

\section{Transport Measurements}
\begin{figure}
\includegraphics[width=4in]{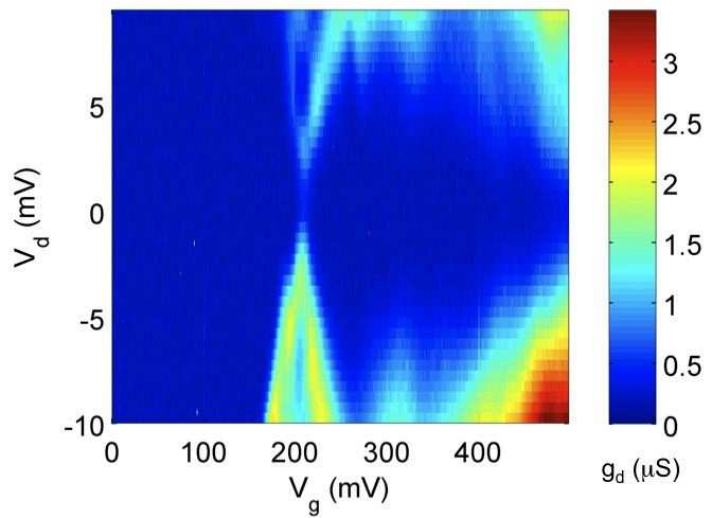}
\caption{Stability diagram of a 28-SOI FinFET ( $l_g$ = 280 nm, $w$ = 120 nm) measured at $T$ = 320 mK. $g_d$ is the conductance of the device. Several overlapping Coulomb diamond features are visible, indicating multiple Coulomb blockade sources in the device.
\label{He3map}}
\end{figure}

\begin{figure}
\includegraphics[width=4in]{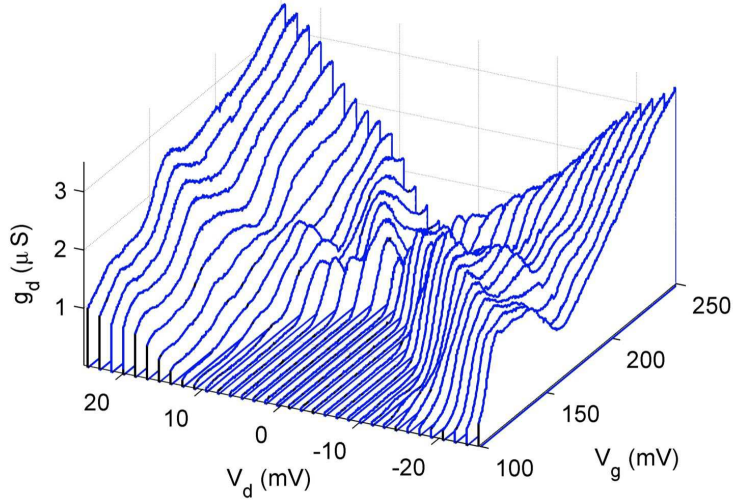}
\caption{Traces of $g_d-V_g$ around the first Coulomb peak at $T$ = 320 mK, revealing oscillatory patterns for $V_d > 0 V$ and a strong resonant feature for $V_d < 0 V$.
\label{He3IdVg}}
\end{figure}

Electrical tranport properties at low temperature are critical for the application of FinFETs as single spin readout devices. The device tested had the SOI layer pre-implanted with $^{121}$Sb at 80 keV with a dose of $6\times10^{11}$ cm$^{-2}$ prior to all other fabrication steps. The post-processing Sb profile peak is expected to be located at 35 nm from the top of the SOI layer, with a peak concentration of $10^{17}$ cm$^{-3}$ from TCAD simulations. The fin width is 80 nm, gate-length 280 nm and height 200 nm. Approximately 130 donor atoms reside in the fin under the gate for the given device dimensions. Low temperature transport measurements were performed with the device mounted in a Helium-3 cryostat with a base temperature of 320 mK. Low-frequency lock-in measurements at 100 Hz and modulation amplitude of 500 $\mu V$ applied to the drain was used to measure the device conductance. The stability diagram of the device at the base temperature is shown in Fig.~\ref{He3map}, with several overlapping Coulomb blockade diamonds visible. The Coulomb blockade structures might be caused by local defects at the oxide-silicon interface or by surface roughness along the channel. It does not appear to be related to quantum confinement under the entire gate length due to the relatively large dimensions of the device. The overlapping diamonds in the stability diagram also indicates independent charge trapping/blockade centers along the conduction path in the device \cite{sohn97}. Fig.~\ref{He3IdVg} shows individual traces of the conductance-gate voltage ($g_d-V_g$) measurements close to the low-voltage corner of the first Coulomb diamond. In the case where the drain voltage is positive, periodic oscillations are observed on the curves. When the drain voltage is biased in the negative regime, a sharp conductance peak is observed at the edge of the diamond edge. The reason for the asymmetry in the transport response is unclear at this point and is under further investigation. 

\begin{figure}
\includegraphics[width=4in]{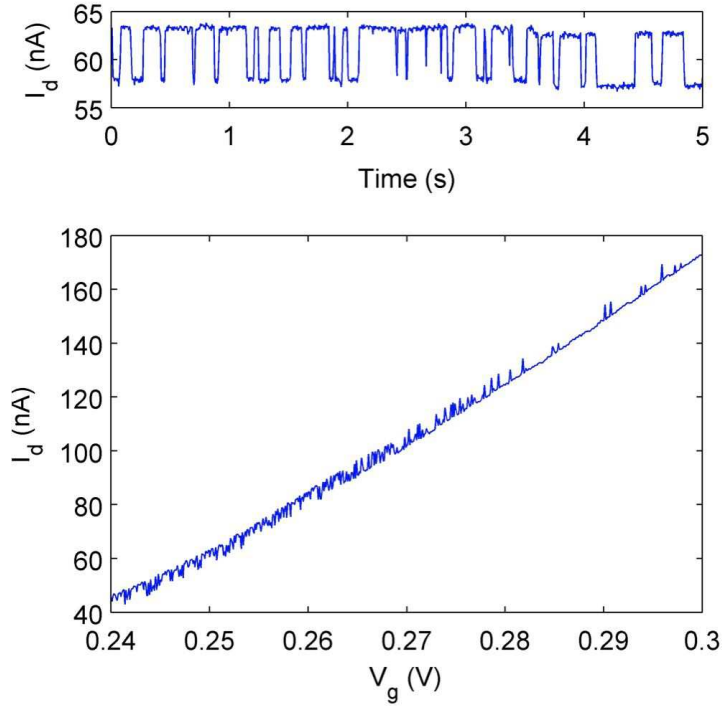}
\caption{ (Upper panel) Random telegraph noise measured at $T \approx $ 10 K with $V_g$ = 0.25 V and $V_d$ = 10 mV for a FinFET with $l_g$ = 280 nm and $w$ = 120 nm. (Lower panel) $I_d-V_g$ measurement of the same device with $V_d$ = 50 mV, the trap state shifts from mostly unoccupied to mostly occupied at around $V_g$ = 0.27V. 
\label{rtn}}
\end{figure}

The device performance is extremely stable \cite{zimmerman07} over $\approx$ 10 hours of measurement time at low temperature ($<$ 1K), and random telegraphic noise is only observable at higher temperatures. Fig.~\ref{rtn} shows one such measurement at T $\approx$ 10 K, revealing the sensitivity of our FinFETs to a single Coulomb scattering center caused by an interface trap. The spin state of a similar trap state has been previously reported in deep sub-micron planer silicon field-effect transistors \cite{xiao04}. A trap state created by the formation of a doubly-occupied shallow donor, the so called D$^-$ center, also perturbs the device current as a Coulomb scattering center. Such D$^-$ centers should then be observable under high magnetic fields and at low temperature\cite{thornton73}, which is a promising candidate as a donor spin-state readout mechanism \cite{mccamey08}.

\section{Conclusions}
We have fabricated accumulation-mode field-effect transistor devices in a FinFET archuitecture using isotopically enriched silicon-on-insulator material (28-SOI) as potential single donor electron spin readout devices.  The FinFET architecture is fully compatible with electrical detection of single-ion implantation events for deterministic single atom doping as well as with single donor spin-state readout via electrically detected magnetic resonance.  Constraints in device design for the FinFETs have been briefly addressed, and low temperature transport measurements show stable device operation over several hours of measurement time.

\ack
We thank AM Tyryshkin and SA Lyon on the ESR measurements of the 28-SOI substrates. This work was supported by the National Security Agency under MOD 713106A, the Department of Energy under Contract No. DE-AC02-05CH11231, the National Science Foundation under Grant No. 0404208, and the Nanoelectronics Research Initiative-Western Institute of Nanoelectronics. Work at the Molecular Foundry was supported by the Office of Science, Office of Basic Energy Sciences, of the U.S. Department of Energy under Contract No. DE-AC02-05CH11231. The Advanced Light Source is supported by the Director, Office of Science, Office of Basic Energy Sciences, of the U.S. Department of Energy under Contract No. DE-AC02-05CH11231. Support in device fabrication by the UC Berkeley Microlab staff is gratefully acknowledged.

\maketitle

\bibliographystyle{unsrt}
\bibliography{lowTpaper}

\maketitle

\end{document}